\documentclass[twocolumn,showpacs,preprintnumbers,superscriptaddress,amsmath,amssymb,prl]{revtex4}

\usepackage{graphicx}
\usepackage{dcolumn}
\usepackage{bm}
\usepackage{amsmath}

\newcommand{\be}{\begin{equation}}
\newcommand{\ee}{\end{equation}}
\newcommand{\ba}{\begin{eqnarray}}
\newcommand{\ea}{\end{eqnarray}}

\begin{document}

\preprint{APS preprint}

\title{Endogenous Versus Exogenous Shocks in Complex Networks: an Empirical Test 
Using Book Sale Ranking}

\author{D. Sornette}
\affiliation{Institute of Geophysics and Planetary Physics
and Department of Earth and Space Sciences,
University of California, Los Angeles, CA 90095}
\affiliation{Laboratoire de Physique de la Mati\`ere Condens\'ee,
CNRS UMR 6622 and Universit\'e de Nice-Sophia Antipolis, 06108
Nice Cedex 2, France}

\author{F. Desch\^atres}
\affiliation{Ecole Normale Sup\'erieure, rue d'Ulm, Paris, France}

\author{T. Gilbert}
\affiliation{Haas School of Business,
University of California, Berkeley, California 94720, USA}

\author{Y. Ageon}
\affiliation{Laboratoire de Physique de la Mati\`ere Condens\'ee,
CNRS UMR 6622 and Universit\'e de Nice-Sophia Antipolis, 06108
Nice Cedex 2, France}

\email{sornette@moho.ess.ucla.edu}

\date{\today}

\begin{abstract}
We study the precursory and recovery
signatures accompanying shocks in complex networks, that we test on
a unique database of the amazon.com ranking of book sales.  
We find clear distinguishing signatures 
classifying two types of sales peaks. Exogenous peaks occur abruptly
and are followed by a power law relaxation, while endogenous
peaks occur after a progressively accelerating power law
growth followed by an approximately
symmetrical power law relaxation which is slower than for exogenous peaks.
These results are rationalized
quantitatively by a simple model of epidemic propagation of
interactions with long memory within a network of acquaintances.
The observed relaxation of sales
implies that the sales dynamics is dominated by cascades
rather than by the direct effects of news or
advertisements, indicating that the social network 
is close to critical. 
\end{abstract}


\maketitle

A fundamental question in the theory of out-of-equilibrium systems is
whether the response function to external kicks can be related to
spontaneous internal fluctuations \cite{ruelle}. At equilibrium, this is
solved by the fluctuation-dissipation theorem connecting susceptibility
and noise. In many complex systems, this question amounts to
distinguishing between endogeneity and exogeneity and is important
for understanding the relative effects of self-organization versus
external impacts. This is difficult in most physical systems because
externally imposed perturbations may lie outside the complex attractor
which itself may exhibit bifurcations. Therefore, observable perturbations
are often misclassified. For this reason,
we have studied a non-physical system in which the dividing line
between endogenous and exogenous shocks is clear in the hope that it
will lead to insights about complex physical systems.  
We study the dynamics of commercial growth and its relaxation in the social system of
interacting buyers, obtained from an Amazon.com database of book sales. 
We do see a characteristic difference in behavior between endogenous and exogenous shocks.

Every book that has sold at least one copy
on the online retailer Amazon is automatically assigned a sales rank. 
Typically, two (respectively ten) sales a day puts a title
in the top 10,000 (respectively 1,000) sellers. The top 100
(respectively 10) sell more than about 30 (respectively 100)
books per day through Amazon. 
Its American website, Amazon.com, updates the
ranks of its top 10,000 books every hour, according to a formula
accounting for recent sales and the entire sales history of the
book. Direct sales are confidential
data but their statistical properties can be reconstructed approximately by
careful observations \cite{Rosenthal}. 
The complementary cumulative distribution $P(s)$
of sales $s$ can be approximated by a stationary
power law  $P(s) =C/s^{\mu}$ with $\mu \approx 2 $ in the range
of sales from a few books sold per day to a few hundreds (see
figure in \cite{Rosenthal}). 
We use this power law to transform
book ranks $r(s)=N P(s)$ into sales $s$ according to the formula
$s = (NC/r)^{1/\mu}$, where $N$ is the total
number of books used to normalize the distribution. Thus, a time
series of the rank $r$ of a given book as a function of time,
sampled at a given rate, can be transformed into a time series of
instantaneous sales flux, through this conversion.

\begin{figure}
\begin{center}
\includegraphics[width=8cm]{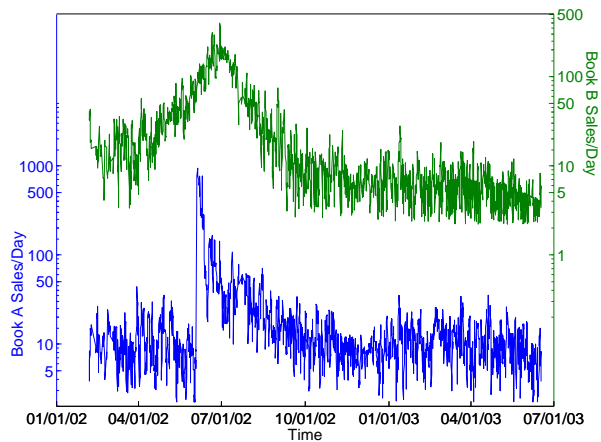}
\end{center}
\caption{ Time evolution over a year and a half of
    the sales per day of two books: Book A (bottom, left scale) 
    is ``Strong Women Stay Young'' by Dr. M.
    Nelson and Book B (top, right scale) is ``Heaven and Earth 
    (Three Sisters Island Trilogy)'' by N. Roberts.}
\label{fig2}
\end{figure}

The time series of ranks of thousands of books have been recorded
with a six-hour sampling rate by JungleScan
(http://www.junglescan.com). From the hundreds of books
with at least one year of recording, we have selected all that have
reached the top 50 in Amazon sales rank. 
We qualify a peak in sales as a local maximum over a three-month time 
window which is at least $k=2.5$ times larger than the average of the 
time series over the three months. In addition, we request that 
there is at least 15 days of data after each peak and 4 days before.
Out of some 14,000 books on Junglescan on April 2004, our algorithm detects
about 1000 such peaks. 
Fig. \ref{fig2} shows about 1.5 years of data for two
books, which are illustrative of the two classes found in this
study. Book A jumped on June 5, 2002, from rank in the 2,000s to
rank 6 in less than 12 hours. On June 4, 2002, the New York Times
published an article crediting the ``groundbreaking research done
by Dr. Miriam Nelson'' and advising the female reader, interested
in having a youthful postmenopausal body, to buy the book and
consult it directly \cite{NYT}. This case is the archetype of an
``exogenous'' shock. 

In contrast, Book B culminated at the end of
June 2002 after a slow and continuous growth, with no such
newspaper article, followed by a similar almost symmetrical decay,
the entire process taking about 4 months. 
The peak for Book B belongs to the class of endogenous shocks as described below.
Qualitatively, such endogenous growth is well explained in Ref.~\cite{tipping} 
by taking the example of the book ``Divine Secrets
of the Ya-Ya Sisterhood'' by R. Wells, which became a bestseller
two years after publication, with no major advertising campaign.
Following the reading of this originally small budget book,
``Women began forming \textit{Ya-Ya} Sisterhood groups of their
own [...]. The word about \textit{Ya-Ya} was spreading [...] from
reading group to reading group, from \textit{Ya-Ya} group to
\textit{Ya-Ya} group'' \cite{tipping}. 

Such social epidemic process can be captured by the following
simple model. The instantaneous sales flux of a
given book results from a combination of external forces such as
news, advertisement, selling campaign, and of social influences in
which each past reader may impregnate other potential readers in
her network of acquaintances with the desire to buy the book. This
impact of a reader onto other readers is not instantaneous as
people react at a variety of time scales. This latency can be
described by a memory kernel $\phi(t-t_i)$ giving the probability
that a buy at time $t_i$ leads to another buy at a later time $t$
by another person in direct contact with the first buyer. Starting
from an initial buyer (the ``mother'' buyer) who notices the book
(either from exogenous news or by chance), she may trigger
buying by first-generation ``daughters,'' who themselves
propagate the buying drive to their own friends, who become
second-generation buyers, and so on. We describe the sum of all buys
by a conditional Poisson branching process with intensity
\be 
\lambda(t)=S(t)+\sum_{i|t_i \leq t} \mu_i~ \phi(t-t_i)~,
\label{lambda}
\ee
where $\mu_i$ is number of 
potential buyers influenced by the buyer $i$ who bought earlier at time $t_i$.
$S(t)$ is the rate of sales initiated spontaneously
without influence from other previous buyers; it can be decomposed
into the sum of a white-noise process with power law distribution
representing small triggering factors (which contribute to the endogenous shocks)
and a jump process (Dirac distributions) modeling massive
media coverage and advertisement campaigns for instance (exogenous shocks).
We note that the distinction between endogenous and exogenous is in general murky
and cannot be decided with 100\% certainty for most books; the correct
approach is probabilistic and relies on the analysis of an ensemble of cases,
as presented below.

Taking the ensemble average of (\ref{lambda}) gives the self-consistent
mean-field equation
\be
s(t)=  \langle \lambda(t) \rangle=
S(t)+ n \int_{-\infty}^{t} d\tau ~\phi(t-\tau)~s(\tau)~,
\label{N1}
\ee 
where $n = \langle \mu \rangle$ is the average number of buys
of first generation triggered by any ``mother'' within her
acquaintance network (also called the branching ratio) and depends on the network topology
as well as on the social behavior of influences.
The Green function $K(t)$ of (\ref{N1})
corresponding to $S(t)=\delta(t)$ (exogenous shock) is easily obtained by taking
its Laplace transform and corresponds to the exogenous response function.
We postulate that
the ``bare propagator'' is of the form $\phi(t-t_i) \sim 1/t^{1+\theta}$ with
$0<\theta<1$ corresponding to a long-memory process that are
commonly observed for sale relaxations. Then, 
\be 
s_{\rm exo} \equiv K(t) \sim 1/(t-t_c)^{1-\theta}~, \label{gmlasa} 
\ee
for $t<t^* \propto 1/(1-n)^{1/\theta}$
and $K(t) \sim  1/t^{1+\theta}$ for $t>t^*$ and $n<1$. Expression (\ref{N1})
can then be written $s(t)=  \int_{-\infty}^{t} d\tau ~K(t-\tau)~S(\tau)$.
Close to the critical
point $n \approx 1$, the cascade of generations embodied in (\ref{N1})
renormalizes the memory kernel $\phi(t-t_i)$ into a
dressed or renormalized memory kernel $K(t-t_i)$
\cite{Endo_Exo}, giving the probability that a buy at time
$t_i$ leads to another buy by another person at a later time $t$
through any possible generation lineage ($\int_{0}^\infty
K(t)dt=n/(1-n)$ is the average number of buys triggered by one buy). 
Thus, if we interpret the
sharp peak of Book A observed in Fig. \ref{fig2} as the impact on
the social network of women created by the extraordinarily favorable
appraisal of the New York Times, the decay of the sales flow that
followed gives a direct measure of its response function $K(t)$:
we find indeed a power law (not shown) as predicted by (\ref{gmlasa})
 with $\theta = 0.3 \pm 0.1$. 
Such power law dependence of the relaxation rate of book sales on Amazon.com
is the hallmark of a long-memory process characterizing the
dynamics of influences within the complex social network. These
laws are similar to the relaxation of seismic activity after an
earthquake, known as Omori's law \cite{Omori} and has also been
found in the response rates of internauts to an exogenous shock,
such as the publication of the URL in a newspaper interview
\cite{www}, and in the relaxation of volatility shocks in the
stock market \cite{Endo_Exo_MRW}. 

In the absence of strong external influences, a peak in book sales
can occur spontaneously due to the interplay between a continuous
stochastic flow of small external news and the amplifying impact
of the epidemic cascade of social influences. We propose that this
mechanism explains the sales time series of books such as Book B
shown in Fig. \ref{fig2}. Technically, the problem amounts to
calculating the average sales trajectory before and after a peak, conditioned
on the existence of a peak. We use the standard result for
stochastic processes $X(t)$ with finite variance and covariance
that $\langle X(t)|X(t=0)=X_0  \rangle \propto {\rm Cov}(X(t),X_0)$.
Applying this result to $\lambda(t)$ defined in (\ref{lambda}) gives
${\rm Cov}(\lambda(t), \lambda_0) \propto \int_{-\infty}^t d\tau ~K(t-\tau)~K(-\tau)$.
This expression gives the average
growth of the sales before such an ``endogenous'' peak and the
relaxation after the peak, which is 
proportional to 
\be
s_{\rm endo}(t) \sim 1/|t-t_c|^{1-2\theta}~,
\label{predictendo}
\ee
for $K(t)$ given by (\ref{gmlasa})
\cite{Endo_Exo}.
The prediction that the relaxation following an
exogenous shock should happen faster (\ref{gmlasa}) (larger exponent $1-\theta$)
than (\ref{predictendo}) for an endogenous shock (with exponent $1-2\theta$) 
reflects the fact that an endogenous shock results from a precursory process that
inpregnates the network much more over a longer time and thus has
a longer lived influence. The prediction (\ref{predictendo})
is verified for Book B with good precision (not shown) 
with the same $\theta = 0.3 \pm 0.1$. 

\begin{figure}
\begin{center}
\includegraphics[width=8cm]{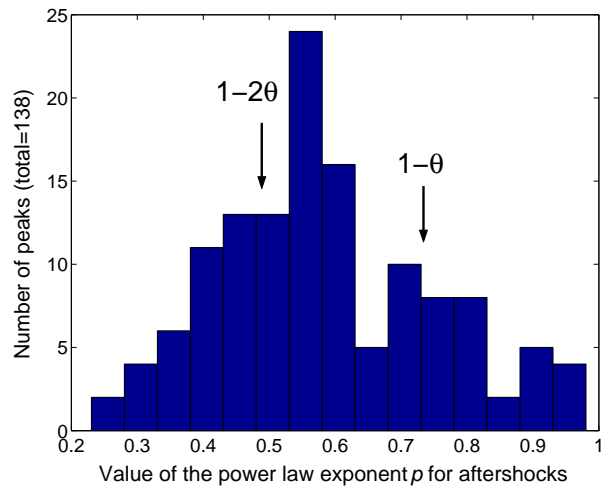}
\end{center}
\caption{\label{fighist} Histogram of the exponents $p$ of the fits
of book sales as a function of time by the power law $\sim 1/(t-t_c)^{p}$.
The arrows indicate the values $p=1-\theta$ (expected for exogenous peaks)
and $p=1-2 \theta$ (for endogenous peaks), with $\theta \approx 0.3$ (see text).
}
\end{figure}

Among the thousand peaks in our database, many are followed by complicated trajectories
probably due to multiple external influences. Nevertheless, we fitted
all sales time dependence after each of the thousand peaks
by a power law of the type (\ref{gmlasa}) or (\ref{predictendo})
and selected for further analysis those that give a correlation coefficient $R$
better than $0.95$. This leaves us with 138 peaks. This criterion ensures that we extract
clear response functions among the generally complex time series. Our purpose
is not so much to show that the response function is a power law but that, 
if it is, then we classify another class of so-called ``endogenous'' peaks with
precursory and decay rates that can be predicted.
Our results below are not changed
qualitatively by changing this threshold on $R$ in the range from $0.8$ to close to $1$.
We also find the good feature that the classification discussed below
in terms of exogenous versus endogenous
peaks improves (fewer misclassifications) as the threshold increases, conforting
the hypothesis that the response function
of a book sales to a strong advertisement is of the form (\ref{gmlasa}).
Our goal is thus 
to extract the pure cases and show that they are consistent with the model, 
which has rigid predictions linking the exponents of the precursory
and relaxation behavior of exogenous and endogenous cases. In turn,
this should permit to use the general formulation (\ref{lambda}) with an arbitrary
time-dependent source term to describe general situations.

For each of the 138 selected peaks, we measure the exponent
$p$ characterizing the power laws $1/(t-t_c)^{p}$ describing the
relaxation of the sales after each peak. 
We perform a mean-square fit from the peak up to a time $t_{\rm end}$, where
$t_{\rm end}$ is varied from 15 days to the first mininum of the sales between
25 days and 6 months after the peak. Once all fits are performed for these 
different time windows, we select the window with the highest correlation coefficient
of the fit to the data
(we also used other criteria such as selecting the window with the smaller
exponent, without altering the results significantly).
The histogram of $p$-values shown in Fig.~\ref{fighist} exhibits
two distinct clusters, one
with a median at $p \approx 0.75$ and the other with a median at $p\approx 0.45$,
compatible with the predictions (\ref{gmlasa}) and (\ref{predictendo})
with the choice $\theta = 0.3 \pm 0.1$. This
classifies the first (respectively second) cluster as exogenous
(respectively endogenous). This histogram is robust with respect
to variations in our procedure, such as different windows and peak thresholds. 

\begin{figure}
\begin{center}
\includegraphics[width=8cm]{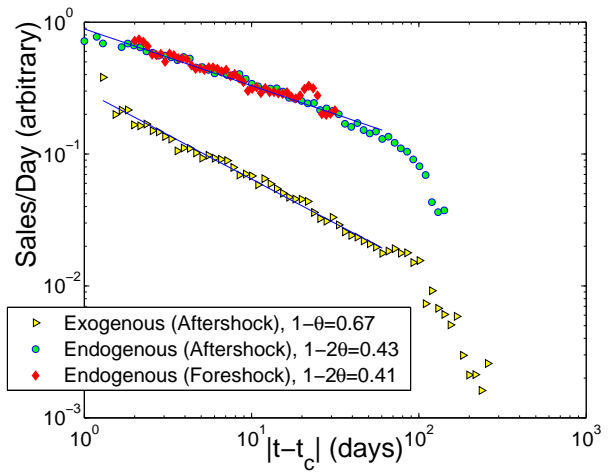}
\end{center}
\caption{Precursory (``foreshock'') and relaxation (``aftershock'') of sales around peaks
obtained after averaging over books in each class classified according
to the precursory acceleration (see text).
}
\label{stack_peak}
\end{figure}

According to our model, the peaks belonging to the cluster with high $p$
($p \approx 0.7$) should be in the exogenous class, and therefore should
be reached by abrupt jumps without detectable
precursory growth. Conversely, the peaks belonging to the cluster
with $p \approx 0.4$ should be in the endogenous class, and therefore
should be associated with a progressive power law precursory growth
$1/(t_{c}-t)^p$ with exponent $p=1-2 \theta$.
To check this prediction, the following algorithm categorizes the growth
of sales before each of the peaks according to its acceleration pattern.
We differentiate between peaks which have an increase in sales in a four day period by a
factor of at least $k_{\mathrm{exo}}$  prior to the peak and peaks that
have an increase in sales by a factor of less than $k_{\mathrm{endo}}$.
We find that the larger $k_{\mathrm{exo}}$ is, the larger is the exponent $p$
of the average relaxation for books that have an increase in sales
by a factor more than $k_{\mathrm{exo}}$. Conversely, the
smaller $k_{\mathrm{endo}}$ is, the smaller is the exponent $p$ of the average
relaxation for books that have an increase in sales by a factor less
than $k_{\mathrm{exo}}$. These results confirm the predictions of the model.
Quantitatively, we first apply a stringent selection
with $k_{\mathrm{exo}} = 30$, $k_{\mathrm{endo}} = 2$. This
implies that peaks, for which the acceleration factor is between $2$ and
$30$, are disgarded in order to get clean signals. 
Out of the $138$ peaks, this leaves us with $30$ peaks. 
We then average the precursory and relaxation behavior of sales for the class of peaks
classified as endogenous with $k_{\mathrm{endo}} = 2$ and as exogenous
with $k_{\mathrm{exo}} = 30$. Fig. \ref{stack_peak} confirms nicely
the existence of two classes, with a symmetric precursory and relaxation behavior
for endogenous peaks, and with all three power laws accounted for 
by a single value $\theta = 0.3 \pm 0.1$. The two clusters
remain significant for less restrictive $k_{\mathrm{exo}}$ and $ k_{\mathrm{endo}}$.
While the theoretical predictions
have been derived for ensemble averages, we find that more than 80\%
among the 138 peaks we have analyzed 
with at least a year of data and that reached the top 50 among all books
also obey $1/(t-t_c)^{1-\theta}$ or $1/(t-t_c)^{1-2\theta}$ individually
with reasonable precision.
Our finding that the classifications of endogenous versus
exogenous match so well on individual books is extremely significant with
only 1 chance in about $10^{8}$ that this result could be obtained by
chance only.

The values of the exponents smaller than one (close to $1-\theta$ and $1-2\theta$)
both for exogenous and endogenous relaxations
imply that the sales dynamics is dominated by cascades involving 
high-order generations rather than by interactions stopping after first-generation 
buy triggering. Indeed, if buys were initiated mostly by the 
direct effects of news or advertisements, and not much by triggering cascades
in the acquaintance network, the cascade model predicts that we should then measure an
exponent $1+\theta$ given by the ``bare'' memory kernel $\phi(t)$.
This implies that the average number $n$ 
of impregnated buyers per initial
buyer in the social epidemic model is on average very close to 
the critical value $1$, because the renormalization from $\phi(t)$ to
$K(t)$ given by (\ref{gmlasa}) only operates close to criticality
characterized by the occurrence of large cascades of buys. 
This offers a new signature of criticality in self-organized networks \cite{Linked}.

Extreme events in complex physical systems, particularly those which seem to involve
self-organized criticality, are often viewed as having an endogenous
source \cite{Bak}.  
Our work shows that the issue is not clear-cut as endogenous and exogenous shocks may lead
to similar power law signatures, which can however be distinguished
by a careful classification. This offers new ideas for probing 
self-organized critical systems as done recently 
for the Olami-Feder-Christensen sandpile model \cite{OFCearth}.
We also note that the distinction between jammed states (constructed by fast
processes) \cite{jammed} versus fragile states 
(formed by slow and delicate accumulation of perturbations) \cite{fragile} of 
granular media and other ``soft-matter'' systems
is based in part on the nature of their preparation and on 
their response to finite and short-lived perturbations
versus infinitesimal continuously repeated ones. Recognizing the
importance of the nature of the perturbation as suggested here could 
provide new insights in the organization of granular media and new 
experimental questions, such as new ways of analyzing the history.
Similar considerations apply to memory retrieval
using hysteresis loops in magnets \cite{olga}.
More generally, physical systems with many competing equilibria such as glasses
and spinglasses are known to betray their history-dependent organization differently 
depending on whether they are subjected to large non-local perturbations versus continuous
slow forcing \cite{Lamarcq}. Our classification of endogenous versus
exogenous shocks presented here should encourage researchers to 
analyze complex physical systems similarly.

{\bf Acknowledgments}: 
We thank W.-X. Zhou and A. Helmstetter for exchanges.
This work was
performed in part while FD and TG were at the Department of Earth and Space
Sciences at UCLA as well as FD at the University of Nice.

\end{document}